\documentclass[aps,prd,notitlepage,showpacs,nofootinbib,superscriptaddress]{revtex4-1}
 \pdfoutput=1
\usepackage[usenames,dvipsnames]{color}  
\usepackage{graphicx}
\usepackage{setspace}
\usepackage{caption}
\usepackage{amsmath}
\usepackage{slashed}
\usepackage{amssymb}
\usepackage[colorlinks=true,linkcolor=darkred,citecolor=darkred,urlcolor=darkred, pdfborder={0 0 0}]{hyperref}
\usepackage[normalem]{ulem}
\usepackage{xcolor}
\usepackage{multirow}
\usepackage{gensymb}
\usepackage{rotating}
\usepackage{braket}
\usepackage{array}
\usepackage{comment}
\usepackage{orcidlink}

\usepackage{booktabs,multirow}
\newcolumntype{C}[1]{>{\centering\arraybackslash}m{#1}}

\makeatletter
\def\p@subsection{}
\makeatother

\definecolor{darkred}{rgb}{0.6,0,0}
\definecolor{denim}{rgb}{0.08, 0.38, 0.74}

\definecolor{linkcolor}{rgb}{0,0,0.5}

\def\gsim{\raise0.3ex\hbox{$\;>$\kern-0.75em\raise-1.1ex\hbox{$\sim\;$}}}
\def\lsim{\raise0.3ex\hbox{$\;<$\kern-0.75em\raise-1.1ex\hbox{$\sim\;$}}}

\def\beqn#1{\begin{equation}\label{#1}}
\def\eeqn{\end{equation}}

\def\beqa#1{\begin{eqnarray}\label{#1}}
\def\eeqa{\end{eqnarray}}

\newcommand {\ignore}[1]{}

\def\lnv{lepton number violation }

\def\321{$\mathrm{SU(3) \otimes SU(2) \otimes U(1)}$ }

\newcommand{\AddrAHEP}{%
  AHEP Group, Institut de F\'{i}sica Corpuscular --
  CSIC/Universitat de Val\`{e}ncia, Parc Cient\'ific de Paterna.\\
 C/ Catedr\'atico Jos\'e Beltr\'an, 2 E-46980 Paterna (Valencia) - SPAIN}
\begin{document}

\title{Probing \lnv at FCC-ee}
\author{Praveen Bharadwaj$\,\orcidlink{0000-0001-7309-101X}\,$}
\email{spspb3454@iacs.res.in}
\affiliation{School of Physical Sciences, Indian Association for the	Cultivation of Science, Jadavpur, Kolkata 700032, India}
\author{Sanjoy Mandal$\,\orcidlink{0000-0003-0171-0752}\,$}\email{smandal@kias.re.kr}
\affiliation{Korea Institute for Advanced Study, Seoul 02455, Korea}
\author{Rojalin Padhan}\email{rojalinpadhan2014@gmail.com}
\affiliation{Department of Physics, Chung-Ang University, Seoul 06974, Korea}
\author{Jos\'{e} W. F. Valle}
\email{valle@ific.uv.es}
\affiliation{\AddrAHEP}
\begin{abstract}
We propose high-multiplicity final-state signatures, such as $e^+e^-\to N\overline{N}\to \ell^+\ell^+ 4j$ with $\ell$ denoting $e,~\mu$, $\tau$,
as probes of lepton number violation (LNV) at FCC-ee, featuring negligible Standard Model background. In contrast to conventional searches such as $pp\to \ell^+ N \to \ell^+ \ell^+ jj$ or the process $e^+e^-\to\nu N$, which are suppressed by the small neutrino masses in conventional seesaw scenarios, the minimal linear seesaw picture avoids this suppression. This enables a direct LNV probe from final-state topology, with over $\mathcal{O}(10^3)$ events expected at FCC-ee. Besides probing the Majorana nature of neutrinos, this offers a novel avenue to test the neutrino mass ordering established by oscillation experiments in a high-energy collider setting. 
\end{abstract}

\maketitle

\section{Preliminaries}

Ever since the discovery of neutrino oscillations~\cite{Kajita:2016cak,McDonald:2016ixn}, a central question in particle physics has been why neutrinos are so light and what underlying symmetries govern the smallness of their masses and the structure of their mixing pattern~\cite{Ding:2024ozt}, as inferred from oscillation data~\cite{deSalas:2020pgw,10.5281/zenodo.4593330,Esteban:2024eli,Capozzi:2025wyn}.
Here we focus on the origin of neutrino masses and its implications. A compelling possibility associates the smallness of neutrino masses with their Majorana nature and the violation of total lepton number. In the classic framework of the seesaw mechanism, however, the mediators responsible for neutrino mass generation are typically super-heavy and therefore inaccessible at colliders.

Nonetheless, the possibility of experimentally probing seesaw mediators at collider experiments~\cite{Dittmar:1989yg,Atre:2009rg,Cottin:2018nms,Alimena:2019zri,Abdullahi:2022jlv,Blondel:2022qqo} has been extensively explored, motivating a variety of genuine low-scale realizations, such as the inverse and linear seesaw models~\cite{Mohapatra:1986bd,GonzalezGarcia:1988rw,Akhmedov:1995ip,Akhmedov:1995vm,Malinsky:2005bi,Batra:2022arl,Batra:2023mds,Batra:2023ssq}, which render such searches feasible. Because lepton number violating (LNV) effects within neutrino oscillations are strongly helicity suppressed~\cite{Schechter:1980gk}, neutrinoless double beta decay remains the most sensitive probe of $\Delta L = 2$ processes~\cite{Schechter:1981bd}.
In a complementary manner, LNV signatures in low-scale seesaw scenarios have also been investigated in collider setups~\cite{Batra:2023mds,Batra:2023ssq}. Here we focus on the associated LNV signatures at the future $\text{FCC-ee}$ involving same-sign dileptons accompanied by four jets.

\section{Large Lepton
Number Violation in linear seesaw model}
Recently there have been several low-scale seesaw studies of collider signatures focusing on the issue of Lepton Number Violation (LNV)~\cite{Anamiati:2016uxp,Antusch:2017ebe,Drewes:2019byd,Fernandez-Martinez:2022gsu,Antusch:2022ceb,Antusch:2023nqd}. 
Here we focus on low-scale seesaw frameworks—including the inverse seesaw and the conventional linear seesaw, in which the lepton number symmetry is broken explicitly but softly. In this case one expects LNV processes to be highly suppressed. This is due to the fact that the contributions from the states forming the quasi-Dirac neutral heavy lepton (NHL) neutrino mass mediators interfere destructively, leading to a strong suppression of LNV signals. 
Nevertheless, if the LNV process is mediated by an on-shell produced heavy neutrinos, oscillations between the two components of the quasi-Dirac pair can occur, potentially spoiling the cancellation that suppresses LNV signals~\cite{Anamiati:2016uxp,Antusch:2017ebe}. The ratio of LNV to LNC event yields can be generally expressed as
\begin{align}
\sigma^{\rm LNV}=\sigma^{\rm LNC} R^{\rm obs} ,  
\end{align}
where $R^{\rm obs}$ takes into account the efficiency of heavy neutrino-antineutrino oscillation in a particular experiment for a specific process~\footnote{ For conventional high-scale seesaw models with heavy Majorana neutrinos $R^{\rm obs}=1$ and hence $N^{\rm LNV}=N^{\rm LNC}$, whereas for quasi-Dirac heavy neutrinos $R^{\rm obs}$ lies anywhere in the range $[0:1]$ depending on the detector geometry and $\Delta M/\Gamma_N$ ratio.}. 

In the high-scale seesaw framework, heavy neutrino production is suppressed by the light–heavy neutrino mixing. As a result, one expects very few lepton-number-conserving (LNC) events, and an even lesser number of LNV events if oscillation effects are negligible~($R^{\rm obs}\ll 1$).  An unsuppressed LNV process would therefore represent a novel feature of broad interest, with both deep conceptual implications and promising prospects for collider searches, providing a viable target for FCC-ee. 
We focus on alternative neutrino mass models with NHLs which feature a non-minimal NHL production mechanism, unsuppressed by light–heavy mixing, such that the LNC production cross section $\sigma(\text{LNC})$ is sizable. As a consequence, the LNV rate $\sigma(\text{LNV})$ can also be enhanced, both in the Majorana case and in the quasi-Dirac scenario, due to the oscillations.
Representative examples of such frameworks include linear seesaw models with an additional Higgs doublet~\cite{Batra:2023ssq,Batra:2022arl,Batra:2023mds}, 
NHL extended leptoquark models~\cite{Padhan:2019dcp,Mandal:2018qpg,Cottin:2021tfo} and  scenarios involving NHL magnetic moment operators~\cite{Chun:2024mus,Beltran:2024twr,Aparici:2009fh}. 

\begin{figure}[]
\centering
\includegraphics[width=0.45\textwidth]{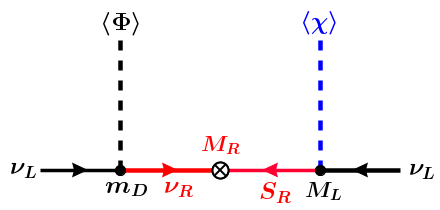}
\caption{\centering
Neutrino mass generation in the linear seesaw mechanism}
\label{fig:numass}
\end{figure}

Motivated by these considerations, here we consider a variant of the minimal linear seesaw model with a single quasi-Dirac NHL mediator. NHL production is governed by a Yukawa coupling ${\bf Y_S}$ which can be sizable while remaining consistent with the smallness of neutrino masses. 
The relevant lepton-number-conserving Lagrangian for neutrino mass generation in this setup is given by~\cite{Batra:2023ssq,Batra:2022arl,Batra:2023mds,Fu:2021fyk}, 
\begin{align}
-\mathcal{L}=Y_\nu^\alpha \bar{L}_\alpha\tilde{\Phi}\nu_R + Y_S^\alpha \bar{L}_\alpha\tilde{\chi}S_R + M_R \overline{\nu_R^c} S_R + \text{H.c.}
\end{align}
with singlets $\nu_R, S_R$ and doublet $\chi$ with lepton number assigned as $L[S_R] = 1 = -L[\nu_R]$ and $L[\chi]=-2$~\cite{Batra:2022arl}. Neutrino mass-generation proceeds as in Fig.~\ref{fig:numass}. This gives the following masses for light and heavy neutrinos, 
\begin{align}\label{lin}
m_{\rm light}\approx \frac{\mathbf{m}_D\mathbf{M}_L^T+\mathbf{M}_L \mathbf{m}_D^T}{M_R}, \,\, \,\,M_{N_{4,5}}=M_R\mp\frac{\Delta M}{2},
\end{align}
with $\mathbf{m}_D=\frac{\mathbf{Y}_\nu v_{\Phi}}{\sqrt{2}}$ and $\mathbf{M}_L=\frac{\mathbf{Y}_S v_{\chi}}{\sqrt{2}}$ where $v_\Phi$, $v_\chi$ denote the vacuum expectation values~(VEVs) of the $\Phi$ and $\chi$ doublets.
Note that, unlike the conventional Type-I seesaw, $m_{\rm light}$ scales linearly with the Dirac Yukawa coupling $\mathbf{Y}_\nu$ contained in $\mathbf{m}_D$, hence the name linear seesaw mechanism. Since only one pair of NHL mediators is introduced in the minimal linear seesaw framework, the overall neutrino mass matrix has rank four, leading to one massless light-neutrino state. The remaining two light-neutrino masses are then fully fixed by the experimentally measured mass-squared differences. 
Notice also that the two heavy states form a quasi-Dirac fermion, so one can define two heavy-neutrino states as 
\begin{equation}
N=(-i N_4 +  N_5)/\sqrt{2},~~~~~\overline{N}=(+ i N_4 +  N_5)/\sqrt{2}.
\end{equation}
where $N$ is always produced in association with an anti-lepton $\ell^+$, while $\overline{N}$ is produced together with a lepton $\ell^-$. This motivates referring to them as $N$ and $\overline{N}$, respectively. As will be discussed later, this definition is particularly useful for describing heavy neutrino-antineutrino oscillations.
\par In this framework, the smallness of the lepton-number–violating term 
$\mathbf{M}_L$ accounts for the lightness of neutrino masses, allowing the scale $M_R$ to remain relatively low despite having sizable ${\bf Y_S}$ Yukawa coupling values. 
This requires a very small $v_\chi$. This can be achieved by soft LNV breaking by the bilinear term $\mu_{12}^2 (\Phi^\dagger \chi + \text{h.c.})$ \footnote{ 
Another scheme with spontaneous LNV is discussed in~\cite{Fontes:2019uld} in which the soft term is replaced by the dynamical VEV of a singlet scalar.}. 
This term induces a nonzero but small VEV for $\chi$, given by
$v_\chi \simeq \mu_{12}^2 v_\Phi / m_A^2$, where $m_A$ denotes the pseudoscalar mass~\footnote{ In the small $v_\chi$ limit, the new scalar spectrum tends to be very much compressed $m_{H^\pm}\approx m_{H,A}$ and within the approximation $m_{H^\pm}>M_N, H^\pm$ dominantly decays to $\ell^\pm N$, see Ref.~\cite{Batra:2023ssq} for details. Subsequently, the NHL decays to SM final states through the light-heavy neutrino mixing $|U|^2$. In this case the mixing between the light and heavy neutrinos is mainly controlled by $U\sim m_D M_R^{-1}$.}. 
In such minimal setup, the mass splitting $\Delta M$ is not arbitrary, but fixed by the splittings actually measured in neutrino oscillation experiments, as $\Delta M^{\rm NO}=\Delta m_{32}$~($\Delta M^{\rm IO}=\Delta m_{21}$). This provides a remarkable connection between the neutrino mass splittings measured in oscillation experiments~\cite{deSalas:2020pgw,10.5281/zenodo.4593330} and the mass splitting of the two heavy neutrino mediators. This feature is key to analyzing the quasi-Dirac nature of heavy neutrino pair in the linear seesaw model.
\par
\begin{figure}
\centering 
\includegraphics[width=0.82\linewidth]{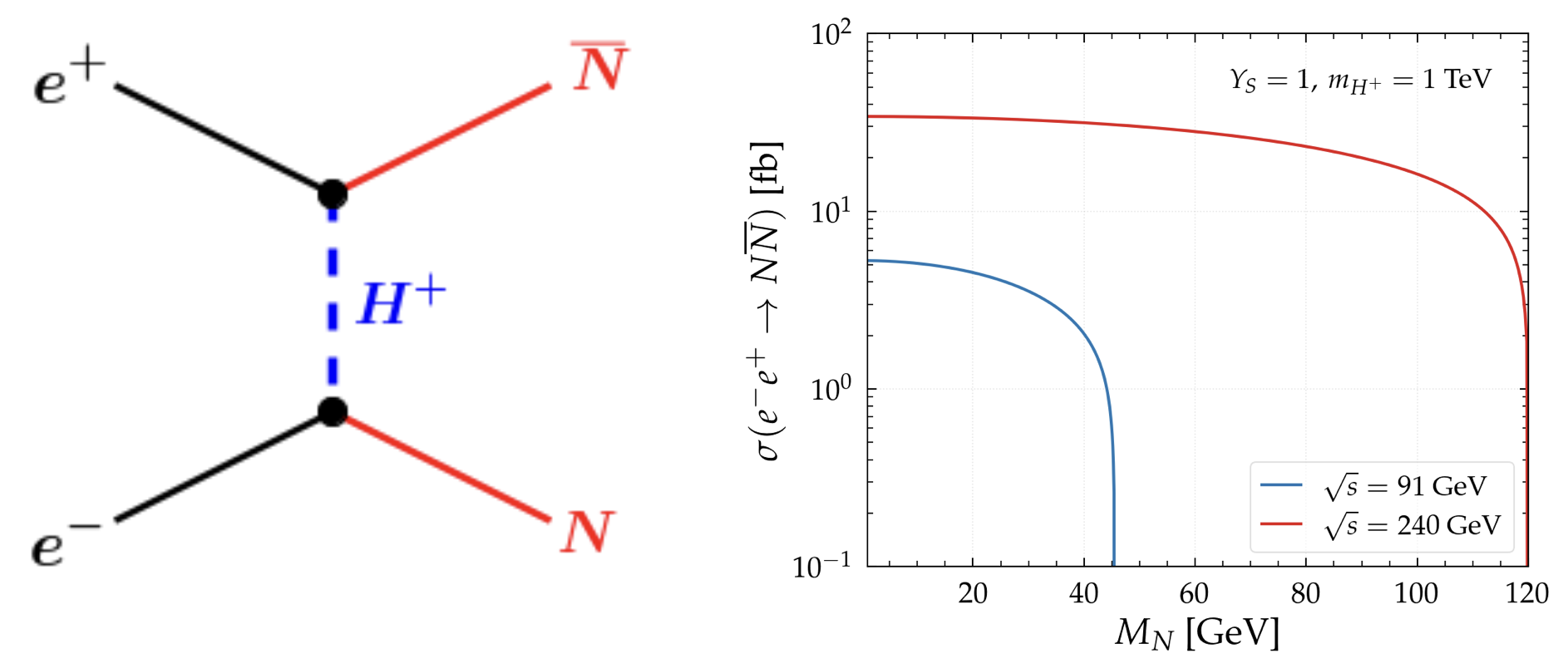}
\caption{ 
Left panel: Feynman diagrams for $e^+e^-\to N\overline{N}$ through t-channel exchange of the charged leptophilic Higgs boson, unsuppressed by light-heavy neutrino mixing. Right Panel: Cross-section as a function of the heavy-neutrino mass, fixing other relevant parameters as ${\bf Y_S}=1$ and $m_{H^+}=1$ TeV.}
\label{fig:cs}
\end{figure}

 A key advantage of the linear seesaw framework, compared to other low-scale realizations such as the inverse seesaw, is that heavy neutrino production is not suppressed by the small light–heavy neutrino mixing.
Indeed, apart from the production through light-heavy neutrino mixing, in our linear seesaw scenario, the heavy neutrino mediators can be produced via the exchange of the leptophilic Higgs boson, controlled by the ${\bf Y_S}$ Yukawa coupling, as illustrated in the left panel of Fig.~\ref{fig:cs}. The contribution to the production cross section arising from the Higgs exchange diagram is proportional to $|{\bf Y_S}|^4$ which can be sizeable for small $v_\chi$ as shown in the right panel of Fig.~\ref{fig:cs}. We show the results for c.m energy $\sqrt{s}=91$ GeV and 240 GeV, whereas we fix other relevant parameters as ${\bf Y_S}=1$ and $m_{H^+}=1$ TeV. Our chosen benchmark is consistent with all existing bounds from collider searches and electroweak precision data (EWPD). 
The small value of $v_\chi$ required by the model, together with EWPD constraints, 
implies that the additional scalar states are nearly degenerate in mass~\cite{Batra:2023mds}. 
Moreover, for such small $v_\chi$, the charged scalar $H^\pm$ decays predominantly into either $\ell^\pm\nu$ or $\ell^\pm N$, depending on whether $m_{H^\pm}<M_N$ or $m_{H^\pm}>M_N$, respectively. 
Consequently, the relevant collider constraints differ in these two mass regimes. 
If the decay $H^\pm\to\ell^\pm\nu$ dominates, current LHC searches impose a lower bound of approximately $m_{H^\pm}\gtrsim700$ GeV~\cite{ATLAS:2019lff,CMS:2020bfa}.  In contrast, when $m_{H^\pm}>M_N$, the dominant decay mode becomes $H^\pm\to\ell^\pm N$, and the corresponding collider limit depends on the subsequent decay channels of the heavy neutral lepton.  Since no dedicated collider search currently exists for this signature, we conservatively adopt the LEP lower bound $m_{H^\pm}>80$ GeV~\cite{ALEPH:2013htx}. Charged lepton flavor violating (cLFV) processes, such as $\mu\to e\gamma$, can impose stringent constraints on the relevant Yukawa couplings. We have verified that the benchmark points considered in this work are consistent with the current cLFV bounds.
\begin{figure}[]
\centering
\includegraphics[width=0.95\textwidth]{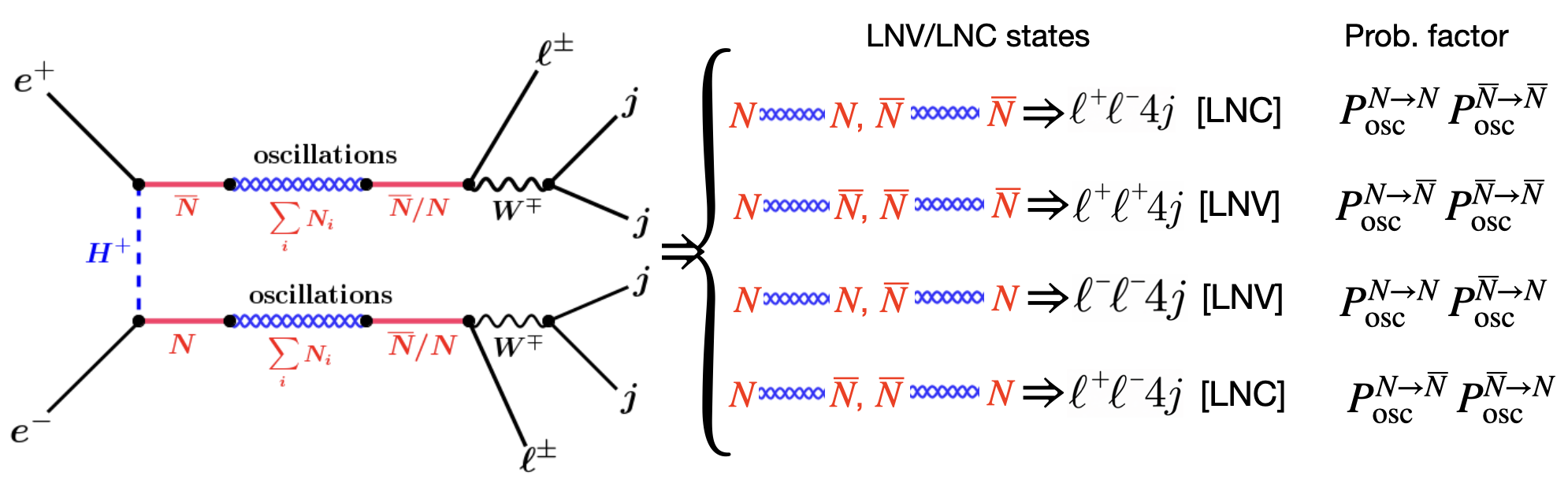}
\caption{
Possible LNV and LNC final states arising from the process $e^+e^-\to\overline{N}N$ followed by NHL oscillations. Due to the coherent superposition of mass eigenstates,
$\sum_i N_i$, the produced heavy neutrinos can oscillate into either
$\overline{N}$ or $N$ before decaying, leading to an antilepton
$(\ell^+)$ or lepton $(\ell^-)$. Possible LNV and LNC signatures along with the associated oscillation probability factors are shown on the right hand side.}
\label{fig:N_osc}
\end{figure}
\par Once produced, the heavy mediators may oscillate before being detected. In contrast to light neutrino-antineutrino oscillations~\cite{Schechter:1980gk}, heavy neutrino-antineutrino conversions are not helicity-suppressed. 
Without such oscillations $N\bar{N}$ pair-production leads only to LNC
 final states, i.e. $e^+ e^- \to N \overline{N} \to ~\ell^\pm \ell^\mp 4j~(\text{LNC})$. 
However, in the presence of such oscillations one also has LNV final states, i.e. $e^+ e^- \to N \overline{N} \to \ell^\pm \ell^\pm 4j~(\text{LNV})$. 
 In other words, in the presence of heavy neutrino oscillation we can have both LNV and LNC final states, as illustrated in Fig.~\ref{fig:N_osc}. The oscillation probabilities of $N(0)\to N(\tau)/\bar{N}(\tau)$ as a function of the proper time $\tau$ are given as $\bar{P}_{\rm osc}^{N\to N(\bar{N})}=e^{-\Gamma_N\tau}/2 \left(1\pm \cos (\Delta M\tau)\right)$~\cite{Batra:2023ssq}. 
The oscillation period is $\tau_{\rm osc}=2\pi/\Delta M$ and the oscillation length is given by $L_{\rm osc}^0=c\tau_{\rm osc}$. 

\par The total probability for an unstable and oscillating particle $N$ to decay in the laboratory frame, 
\begin{align} 
P_{\rm osc}^{N\to N(\bar{N})}(x) =\frac{1}{2L_N}\int_{x_1}^{x_2}e^{-x/L_N}\left(1\pm \cos (2\pi x/L_{\rm osc})\right) dx,
\end{align}
where $L_N=L_N^0\sqrt{\gamma^2-1}$ and $L_{\rm osc}=L_{\rm osc}^0\sqrt{\gamma^2-1}$ are the heavy neutrino decay length and oscillation length in the laboratory frame, with $\gamma=E_N/M_N$ denoting the Lorentz factor for the specific process. 
In our minimal linear seesaw scheme, the mass-splitting is $\Delta M=41.51\times 10^{-3}$ eV for normal mass ordering~(\textbf{NO}), and $\Delta M=749.8\times 10^{-6}$ eV for inverted ordering~(\textbf{IO}). 
The corresponding oscillation lengths~($L_{\rm osc}^0$) are $2.98\times 10^{-5}$~m and $1.65\times 10^{-3}$~m, respectively. 
Notice that if the Lorentz factor is large, the oscillation length in the laboratory frame can be large enough to be experimentally resolvable at FCC-ee~\cite{Antusch:2017ebe}, particularly for {\bf IO}.

The expected number of heavy neutrinos produced at the interaction point and decaying with a displacement of at least $x_1$ and at most $x_2$, can be obtained by combining the production and decay of heavy neutrinos~\footnote{  The vertex displacement $x$ is defined as the distance between the primary vertex where the heavy neutrino~($N$ or $\bar{N}$) with finite lifetime was produced, and its secondary decay vertex. 
}. 
\begin{figure}
\centering
\includegraphics[width=0.45\linewidth]{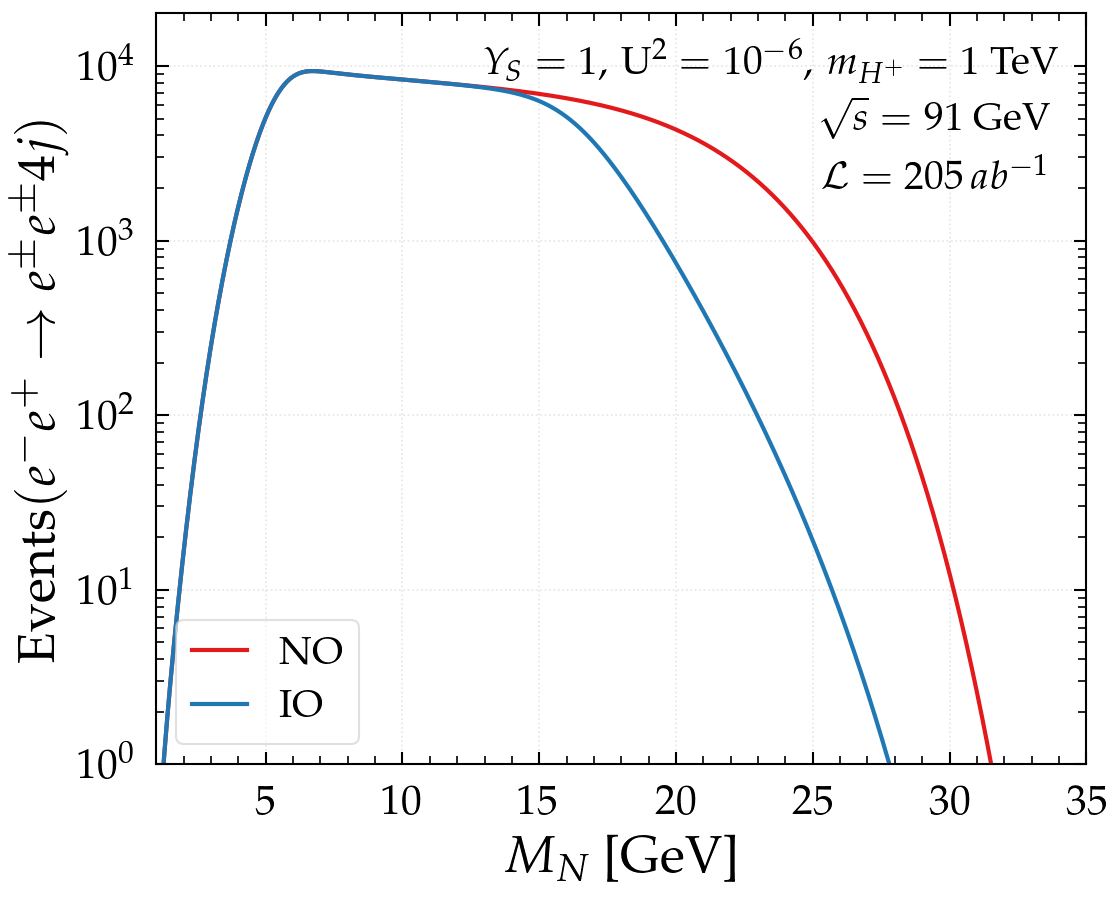}
\includegraphics[width=0.45\linewidth]{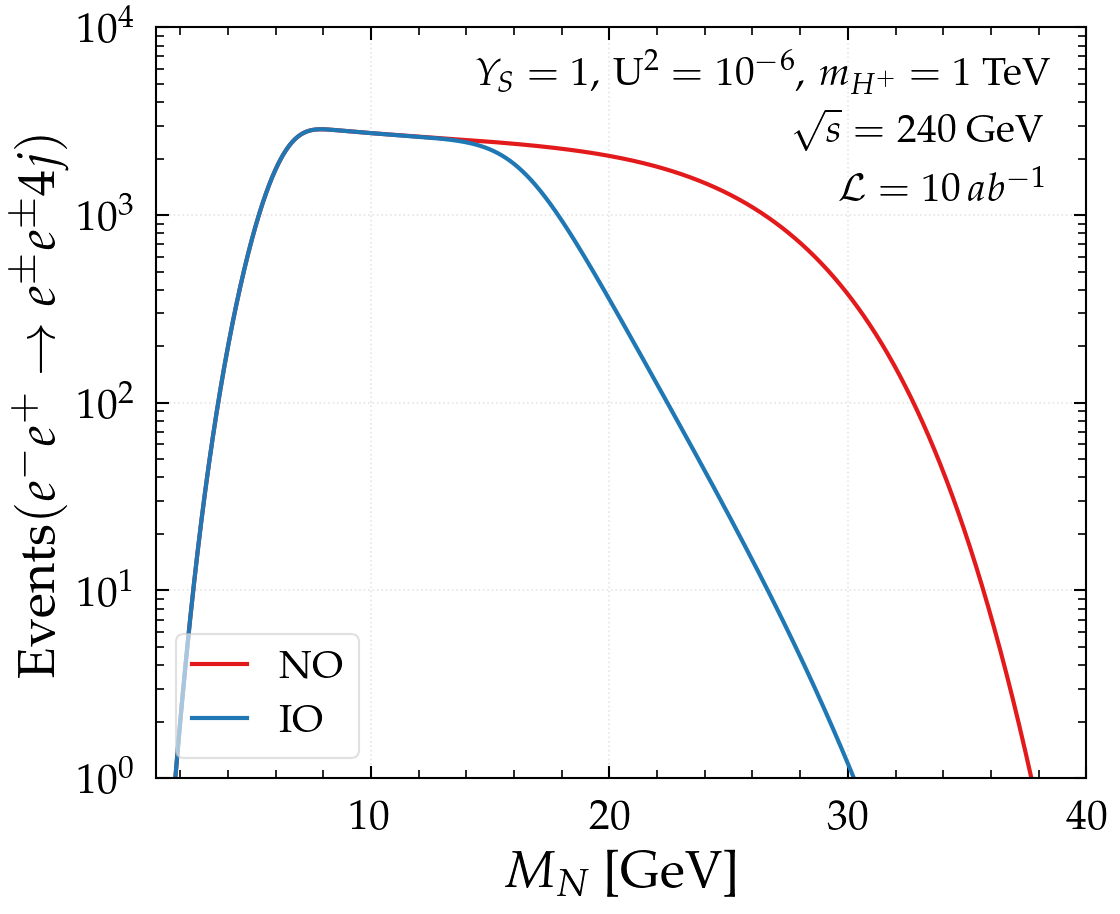}
\caption{ 
Expected number of LNV events at FCC-ee with c.m energy $\sqrt{s}=91$ GeV and 240 GeV, with luminosities $205\text{ ab}^{-1}$ and $10\text{ ab}^{-1}$, respectively. We fix the light-heavy neutrino mixing as $U^2=10^{-6}$ where $U^2=\sum_{\alpha}|U_{\alpha}|^2$. Similar results can be obtained for muon final states.}
\label{fig:events}
\end{figure}
Hence, the corresponding number of LNC and LNV events is, 
\begin{align}  
& N^{\rm LNC}(x_1,x_2,\sqrt{s},\mathcal{L})=\mathcal{L}\, \sigma\, \text{BR} \, \Big[\big(P_{\rm osc}^{N\to N}(x_1,x_2)\big)^2 + \big(P_{\rm osc}^{N\to \overline{N}}(x_1,x_2)\big)^2 \Big], \\
& N^{\rm LNV}(x_1,x_2,\sqrt{s},\mathcal{L})=2\mathcal{L}\, \sigma\, \text{BR} \, P_{\rm osc}^{N\to N}(x_1,x_2) P_{\rm osc}^{N\to \overline{N}}(x_1,x_2),
\label{Eq:LNV}
\end{align}
where $\mathcal{L}$ is the luminosity of the collider, $\sigma$ and BR are the heavy neutrino production cross-section and relevant decay branching ratios. In deriving the above expressions, we use the following relations, $P_{\rm osc}^{\overline{N}\to \overline{N}}=P_{\rm osc}^{N\to N}$ and $P_{\rm osc}^{\overline{N}\to N}=P_{\rm osc}^{N\to \overline{N}}$, where we assumed CP invariance. We have neglected such effects in our analysis. Note that the factor 2 in Eq.~\ref{Eq:LNV} takes into account both possible LNV final states $\ell^+\ell^+$ and $\ell^-\ell^-$, where $\ell$ denotes $e,~\mu$ or $\tau$.
\par If the detector size is infinite, a large number of oscillations take place inside the detector volume, the oscillation probability averages out as 
\begin{align}
 P_{\rm osc}^{N\to N}(0,\infty) = \frac{1}{2}\left(
\frac{\Gamma_N^2}{\Delta M^2 + \Gamma_N^2} + 1\right), \text{  and  }
P_{\rm osc}^{N\to \bar{N}}(0,\infty) = \frac{1}{2}
\frac{\Delta M^2}{\Delta M^2 + \Gamma_N^2}.
\end{align}
Hence, we see that the oscillation probability depends on the ratio of $\Delta M$ to $\Gamma_N$. Oscillations will be effective as long as $\Delta M > \Gamma_N$. In this case the ratio between the LNC and LNV events is given by
\begin{align}
R=\frac{N^{\rm LNV}}{N^{\rm LNC}}=\frac{y^2(2+y^2)}{2+y^2(2+y^2)} \text{  with  } y=\frac{\Delta M}{\Gamma_N}.
\end{align}
When the mass splitting $\Delta M$ is larger than a few times the decay width $\Gamma_N$~($y\gtrsim 4$), one finds that $R$ approaches rapidly the limit $R=1$ see for example Fig.~9 of Ref.~\cite{Batra:2023ssq}. Since the decay width scales as $\Gamma_N\propto |U|^2$, one can easily achieve the condition $y\gtrsim 4$ as long as mixing is relatively small. This leads to a remarkable result, namely, since that the mass splitting $\Delta M$ depends on the neutrino mass ordering, the NHL oscillation—and consequently the LNV event rates—also exhibit sensitivity to the neutrino mass ordering. 
\par Notice that, even if the parameters do not allow for the heavy-neutrino oscillations to be
resolvable, an integrated effect could still be measured. 
We now  discuss this
integrated effect on the expected number of LNV and LNC events for FCC-ee. A complete phenomenological analysis should incorporate the time evolution of both the luminosity and the center-of-mass energy over the full operational period of the FCC-ee. As a first approximation here we simply take two fixed center-of-mass energies $\sqrt{s}=91$~GeV and $\sqrt{s}=240$ GeV with the total integrated luminosities $\mathcal{L}=205\text{ ab}^{-1}$ and $\mathcal{L}=10\text{ ab}^{-1}$, respectively~\cite{FCC:2025lpp}. We assume the vertex displacement to lie within the range $1\text{ mm}\leq x \leq 1\text{ m}$~\footnote{ 
For FCC-ee, we can estimate the total number of heavy neutrino decays inside a cylindrical detector of length $l_{\rm cyl}$ and diameter $d_{\rm cyl}$ by  setting $x_2=(1/2)\, (3/2)^{1/3} d_{\rm cyl}^{2/3}l_{\rm cyl}^{1/3}$, so that a sphere of radius $x_2$ has the same volume as the cylinder~\cite{Drewes:2022rsk}. We find $x_2\approx \mathcal{O}(1\,\text{m})$ and, for definiteness, take the maximum displacement as $x=1\,\text{m}$.}.
In Fig.~\ref{fig:events} we present the expected number of lepton-number-violating (LNV) events~($e^+ e^-\to e^\pm e^\pm 4j$) events at FCC-ee for center-of-mass energies $\sqrt{s}=91$~GeV~(left panel) and $\sqrt{s}=240$~GeV~(right panel), shown as a function of the NHL mass $M_N$ for both for both {\bf NO} and {\bf IO} cases. 
We fix the active–sterile mixing to be $U^2=10^{-6}$, the Yukawa coupling to $|{\bf Y_S}|=1$ and the charged Higgs mass to $m_{H^\pm}=1$~TeV. 

One sees that the number of LNV events decreases sharply above a certain value
of $M_N$, as heavy neutrino oscillations become ineffective when the mass splitting is smaller than the decay width, i.e.~$\Delta M <\Gamma_N$. 
This suppression occurs at a lower
$M_N$ value in the {\bf IO} case compared to {\bf NO}, due to the fact that $\Delta M^{{\bf NO}} > \Delta M^{{\bf IO}}$, so that $P_{\rm osc}^{{\bf NO}}>P_{\rm osc}^{{\bf IO}}$ for fixed mixing parameter $U$ and mass $M_N$. 
Notice also that
no events are observed at low $M_N$, because the NHL lifetime becomes sufficiently long that its decay length exceeds the detector size, leading to decays predominantly outside the detector volume.
To sum up, within the linear seesaw framework, there exists a broad parameter region where both NHL production and oscillation effects are significant, yielding LNV event rates as large as
$\mathcal{O}(10^3)$ at FCC-ee.

\section{Summary and outlook}

A key advantage of our linear seesaw framework is that heavy neutrino production is not suppressed by the smallness of neutrino masses, as it is governed by the Yukawa coupling $\mathbf{Y}_S$ rather than light–heavy mixing.  
This allows for sizable LNV rates even in the regime $\Delta M < \Gamma_N$. The signal features high-multiplicity final states, e.g. $e^+e^-\to N\bar{N}\to \ell^+\ell^+ 4j$, with negligible irreducible SM background, i.e. no SM process yields this final state at tree level. Reducible backgrounds such as those arising from charge misidentification, non-prompt leptons, and jet mismeasurement are instrumental in origin and are left to a future dedicated study. This setup enables direct probes of the Majorana nature from final-state topology. This is in contrast to conventional studies relying on $pp\to \ell^+ N \to \ell^+ \ell^+ jj$ which is neutrino-mass-suppressed, through the 
light–heavy neutrino mixing factor. Likewise, the process $e^+e^-\to\nu N$ suffers, in addition,  from missing energy measurements.
We note also that, in linear seesaw schemes, the presence of fundamental Majorana CP violating phases~\cite{Schechter:1980gr} could also lead to other observable effects~\cite{Abada:2019bac,Godbole:2020jqw}.
Finally, one can also show~\cite{Batra:2023ssq} that
the simple relation between the heavy neutrino mass splitting and neutrino oscillation parameters opens up the possibility to test the neutrino mass ordering at high-energy collider experiments, as shown in~\cite{Batra:2023ssq}.

\section*{Acknowledgements}

The authors sincerely thank Patrick Janot for proposing this analysis and for suggesting the benchmark scenarios and center-of-mass energies currently envisioned for the FCC-ee. This work is funded by Spanish grant PID2023-147306NB-I00 and by Severo Ochoa Excellence grant CEX2023-001292-S (MCIU/AEI/10.13039/501100011033), and also by Generalitat Valenciana: Prometeo CIPROM/2021/054.
The work of S.M. is supported by KIAS Individual Grants (PG086002) at the Korea Institute for Advanced Study.
P.B. acknowledges the financial support provided by the Indian Association for the Cultivation of Science, Kolkata.
The work of RP is supported in part by Basic Science Research Program through the National
Research Foundation of Korea (NRF) funded by the Ministry of Education, Science and
Technology (NRF-2022R1A2C2003567). 

\bibliographystyle{utphys}
\bibliography{bibliography.bib}
\end{document}